\documentclass[a4paper,12pt]{article}
\usepackage[margin=1.3cm]{geometry}
\usepackage{comment}
\usepackage{amssymb,extarrows,graphicx,subfigure,setspace}
\usepackage{cite}
\usepackage{slashed}
\usepackage{tensor}
\usepackage[toc,page]{appendix}
\usepackage{color}
\usepackage{physics}
\usepackage{hyperref}
\usepackage{dirtytalk}
\hypersetup{colorlinks=true, linkcolor=blue, citecolor=red, linktoc=page}
\makeatother

\usepackage{float}
\usepackage{textcomp}
\usepackage{amsmath}
\newmuskip\pFqmuskip

\newcommand*\pFq[6][8]{%
  \begingroup 
  \pFqmuskip=#1mu\relax
  \mathcode`\,=\string"8000
  \begingroup\lccode`\~=`\,
  \lowercase{\endgroup\let~}\pFqcomma
  {}_{#2}F_{#3}{\left[\genfrac..{0pt}{}{#4}{#5};#6\right]}%
  \endgroup
}
\newcommand{\pFqcomma}{\mskip\pFqmuskip}

\usepackage{mathrsfs}
\usepackage{hyperref}

\newcommand{\be}{\begin{equation}}
\newcommand{\bea}{\begin{eqnarray}}
\newcommand{\eea}{\end{eqnarray}}
\newcommand{\ba}{\begin{array}}
\newcommand{\ea}{\end{array}}
\newcommand{\ee}{\end{equation}}
\newcommand{\bes}{\begin{equation*}}
\newcommand{\beas}{\begin{eqnarray*}}
\newcommand{\eeas}{\end{eqnarray*}}
\newcommand{\bas}{\begin{array*}}
\newcommand{\eas}{\end{array*}}
\newcommand{\ees}{\end{equation*}}

\numberwithin{equation}{section}

\begin{document}
\color{black}
\begin{center}
\Large{\bf Weak Cosmic Censorship and Weak Gravity Conjectures\\ in CFT Thermodynamics}\\
\small \vspace{1cm}
{\bf Jafar Sadeghi$^{\star}$\footnote {Email:~~~pouriya@ipm.ir}}, \quad
{\bf Saeed Noori Gashti $^{\dag}$\footnote {Email:~~~saeed.noorigashti@stu.umz.ac.ir}},\quad
{\bf Mohammad Reza Alipour$^{\dag,\star}$\footnote {Email:~~~mr.alipour@stu.umz.ac.ir}},\quad\\\vspace{0.2cm}
{\bf Mohammad Ali S. Afshar $^{\star}$\footnote {Email:~~~m.a.s.afshar@gmail.com}},\\
\vspace{0.3cm}$^{\star}${Department of Physics, Faculty of Basic
Sciences,
University of Mazandaran\\
P. O. Box 47416-95447, Babolsar, Iran}\\
\vspace{0.3cm}$^{\dag}${School of Physics, Damghan University, P. O. Box 3671641167, Damghan, Iran}
\small \vspace{0.3cm}
\end{center}
\begin{abstract}
In this paper, we explore the intriguing interplay between fundamental theoretical physics concepts within the context of charged black holes. Specifically, we focus on the consistency of the weak gravity conjecture (WGC) and weak cosmic censorship conjecture (WCCC) in the thermodynamics of conformal field theory (CFT), and restricted phase space thermodynamics (RPST) for AdS Reissner-Nordström black holes with a perfect fluid dark matter (RN-PFDM). The WGC ensures that gravity remains the weakest force in the system. Meanwhile, the WCCC addresses the cosmic censorship problem by preventing the violation of fundamental physical laws near the black hole singularity. First, we analyze the RN black hole's free energy in both spaces, revealing a distinctive swallowtail pattern indicative of a first-order phase transition when certain free parameter conditions are met. We explore the  WGC across different phase spaces, emphasizing the need for certain conditions in extended phase space thermodynamics (EPST), RPST, and CFT. We demonstrate that PFDM parameter \( \gamma \) and the radius of AdS \( l \) have a vital role in proving the satisfaction of the WGC. Also, these values have a linear relation with the range compatibility of WGC. The range of compatibility for WGC in RPST and EPST is the same while for CFT, this range is larger than EPST, and RPST. It means somehow the WGC and CFT are more consistent. The WCCC was examined at the critical juncture, confirming its validity in critical points. We conclude that the WGC is supported at the critical point of black holes, and the WCCC is also maintained, demonstrating the robustness of these conjectures within the critical ranges of black hole parameters.
\end{abstract}
\tableofcontents
\section{Introduction}
The principles of holography and the AdS/CFT duality are instrumental in examining various black hole properties, where a state of an AdS black hole in the bulk corresponds to a thermal state in the dual field theory\cite{1,2,3,4,5}. The thermodynamics of black holes can be analyzed through the lens of different symmetries within a gravitational model\cite{6}. Hawking and Page's advanced research on AdS black holes revealed a phase transition between such black holes accompanied by radiation and their thermal AdS counterparts\cite{7}. The thermodynamic properties of black holes, especially within the framework of extended phase space thermodynamics, have been extensively discussed, where the negative cosmological constant is equated with pressure\cite{8,9,10,11,12}. This perspective has led to the inclusion of other model parameters as novel thermodynamic variables\cite{13,14,15,16}.\\\\
A recent theoretical development has introduced central charge and chemical potential as a new pair of conjugate thermodynamic variables, enhancing our understanding of black hole thermodynamics. In the corresponding dual theory, the central charge is linked to the square of the number of colors\cite{17,18,19}.
In the realm of AdS/CFT correspondence, a novel thermodynamic framework known as RPS has emerged, as introduced by Gao et al\cite{21,1200}. This paradigm redefines black hole masses as a form of internal energy. Unlike the EPST which utilizes pressure and volume, RPST employs a distinct set of conjugate thermodynamic variables: the chemical potential, denoted as $\mu$, and the central charge, C. These variables represent color susceptibility and the degrees of freedom within the conformal field theory (CFT), respectively\cite{18,19,20}.
The central charge, C, plays a role analogous to the particle count in the statistical physics of conventional matter. It is mathematically expressed in relation to Newton's constant, G, and the anti-de Sitter (AdS) curvature radius, through the formula $ C = \frac{l^{d-2}}{G} $. Here, $ d $ signifies the dimensionality of the AdS spacetime. With respect to demonstrating that altering G alongside \(\Lambda\) maintains the constancy of the central charge C in the dual CFT, so can consequently, the first law can be reformulated to include a new thermodynamic volume and chemical potential, bridging holographic and bulk thermodynamics through the central charge duality. The RPST framework has been applied to explore the homogeneity of the Smarr relation, various thermodynamic processes, and phase transitions across different black holes\cite{20,21,22,23,24,25}.\\\\
The properties and phase transitions of black holes in AdS space through holographic duality provide insights into phenomena such as the Hawking-Page transition, which reflects a phase shift from thermal AdS space to a large AdS black hole, analogous to the confinement/deconfinement transition in conformal field theory. One of the intriguing outcomes is that AdS black holes adhere to a generalized Smarr relation, linking their mass, area, charge, angular momentum, and pressure. However, integrating the first law of black hole thermodynamics with the thermodynamics of its holographic dual presents challenges. Researchers are particularly focused on how the variation of the cosmological constant (\(\Lambda\)) correlates with changes in the central charge (C) and the volume (V) of the boundary CFT\cite{26,27,28,29}. They also uncovered phase structures with a critical central charge value, beyond which the phase diagram shows a swallowtail pattern indicative of a first-order phase transition. By treating G as a variable, we can incorporate the central charge into the first law and extend the thermodynamic discussion to various black holes. Yet, the extended thermodynamics' holographic interpretation remained elusive for some time. Initial theories proposed that the volume-pressure term in AdS should correspond to a chemical potential-central charge term in the dual CFT, where the central charge is C and the chemical potential is its thermodynamic conjugate\cite{26,27,28,29}.\\\\
The swampland program is an initiative in theoretical physics that seeks to identify effective low-energy theories that are compatible with quantum gravity. The swampland encompasses theories that appear consistent but lack a consistent ultraviolet completion when gravity is added\cite{30,31,32,33,34,35,36,37,38}. The Weak Gravity Conjecture is a theoretical framework in the field of quantum gravity that suggests gravity should be the weakest force in any consistent theory of quantum gravity. It posits that for any gauge force, there must exist some particles for which the gauge force mediates interactions stronger than gravity\cite{30,31,32,33,34}. The WGC is not without its criticisms and potential weak points. Some of the criticisms include that the conjecture is based on qualitative arguments rather than a rigorous mathematical proof \cite{o}.
The WGC has made several predictions, including:
The existence of particles with specific mass-to-charge ratios ensures gravity remains the weakest force\cite{30,31,32,33,34}.
Constraints on the properties of black holes, particularly regarding their charge and mass.
Implications for the stability of cosmic structures and the behavior of fundamental particles\cite{30,31,32,33,34}.\\
The WGC has interesting connections with thermodynamics, particularly in the context of black hole physics. It has been suggested that the conjecture could be derived from thermodynamic principles applied to black holes, such as entropy considerations and the behavior of quasinormal modes \cite{ee,hh}. The implications of the WGC can extend to various spacetime geometries, including extended, restricted, and CFT spaces:
The conjecture has been explored in the context of theories with extra dimensions, such as string theory, where it places constraints on the geometry of the compactified dimensions\cite{30,31,32,33,34}.
In more constrained spacetime geometries, like those with a limited number of dimensions or symmetries, the WGC still implies the existence of particles that satisfy its mass-to-charge ratio condition\cite{30,31,32,33,34}.
In the AdS/CFT correspondence, the WGC translates into inequalities involving the dimension and charge of operators, as well as central charges in the dual CFTs. However, these translations are still speculative and may not apply universally across all CFTs \cite{oooo,ooooo}. The WGC has far-reaching implications that extend beyond the realm of theoretical physics into cosmology, black hole thermodynamics, and the broader landscape of string theory. In cosmology, the WGC can influence our understanding of the early universe, particularly during the inflationary period. The conjecture suggests that inflationary models involving scalar fields should be coupled to gauge fields in a way that satisfies the WGC criteria \cite{oooooo}. This can lead to predictions about the spectrum of primordial fluctuations, potentially observable in the Cosmic Microwave Background (CMB) data. String theory provides a fertile ground for testing the WGC. The conjecture has been used to argue for the existence of certain particles in the string spectrum that have implications for the stability of extra dimensions and the nature of supersymmetry breaking. A proof of the WGC within the context of perturbative string theory has been proposed, which relates the black hole extremality bound to long-range forces calculated on the worldsheet \cite{aaaaa}. The thermodynamics of black holes offers a unique testing ground for the WGC. The conjecture has been studied in the context of 4D Einstein-Maxwell-dilaton theory, where it places constraints on the Wilson coefficients of the theory, which are satisfied under certain assumptions about UV completion. This study bridges the gap between quantum field theory and gravitational phenomena, providing insights into the entropy and stability of black holes \cite{aaaa}. The WGC may serve as a bridge between quantum mechanics and cosmic-scale physics. By examining the thermodynamics of black holes and their connection with the WGC, we can explore how quantum-level interactions manifest at cosmic scales. This could lead to a better understanding of the holographic principle and the nature of spacetime itself.
The phenomenological implications of the WGC are vast. It can affect predictions for particle accelerators, inform the search for dark matter candidates, and influence models of dark energy. The conjecture's requirement for the existence of certain particles could lead to observable signatures in high-energy physics experiments. For more study, you can see in\cite{c,d,e,f,g,h,i,j,k,l,m,n,n',n'',p,p',q,r,s,t,u,v,w,x,y,z,aa,bb,cc,dd,ff,gg,ii,jj,kk,ll,mm,nn,oo,pp,qq,rr,ss,tt,uu,vv,ww,xx,yy,zz,aaa,bbb,ccc,ddd,eee,fff,ggg,hhh,iii}.\\\\
In this exploration, we examine the nuanced interplay between the  WGC and the WCCC in the realm of charged black holes. Our focus is on the (RN) black holes that are encased within PFDM, considering diverse spatial frameworks like RPS and CFT. This analysis aims to deepen our understanding of the theoretical underpinnings that govern these cosmic entities. The WCCC posits that singularities resulting from gravitational collapse are invariably concealed behind event horizons, thereby maintaining the causal structure and predictability of spacetime. This conjecture ensures that the "naked" singularities, which could expose the inner workings of black holes, do not disrupt the observable universe. In contrast, the WGC suggests that in any consistent theory of quantum gravity, there must exist particles for which the electromagnetic force is stronger than their gravitational pull. This implies that black holes should be able to decay into these particles, preventing the formation of stable remnants and ensuring the discharge of black holes. The Reissner–Nordström (RN) black hole, a solution to the Einstein-Maxwell equations for a charged black hole, presents a unique challenge to these conjectures. If the charge \( Q \) of the black hole exceeds its mass \( M \) (\( Q > M \)), the RN solution would contravene the WCCC by revealing a naked singularity to distant observers. Conversely, when the RN black hole approaches an extremal state (\( Q = M \)), it suggests the existence of decay products with a charge-to-mass ratio exceeding unity. While this satisfies the WGC, it simultaneously breaches the WCCC, as these decay products cannot form black holes but are instead predicted to be elementary particles. This apparent contradiction between the WGC and the WCCC in the context of RN black holes is one of the intriguing puzzles at the intersection of quantum theory and gravitational physics\cite{39,40,41}.\\\\
The  WGC and the WCCC are two distinct ideas in theoretical physics, each addressing different aspects of fundamental physics\cite{999,9999,99999}. While both conjectures involve the term "weak" and pertain to fundamental aspects of gravity and spacetime, they are distinct in their focus and implications. The WGC primarily concerns the relative strengths of forces in a quantum gravity framework, while the WCCC deals with the behavior of singularities and event horizons in gravitational collapse scenarios within classical general relativity. Combining these concepts, one might speculate about potential connections between the WGC and cosmic censorship, such as whether the WGC could have implications for cosmic censorship scenarios. In\cite{45555} researchers explore how Reissner-Nordström (R-N) black holes in perfect fluid dark matter (PFDM) might reconcile the weak gravity conjecture (WGC) and the cosmic censorship conjecture (WCCC). Normally, a charged black hole cannot violate the WCCC, but recent proposals suggest a connection between the WGC and WCCC. The study shows that with certain constraints, both conjectures can coexist. Without PFDM, an R-N black hole can lead to a naked singularity, violating the WCCC. However, with PFDM, the black hole always has event horizons, covering the singularity and satisfying both conjectures. A critical value of PFDM makes the black hole extreme, still fulfilling both WGC and WCCC. Thus, PFDM allows compatibility between the WGC and WCCC in R-N black holes. In this work, we intend to consider another approach to investigate the weak gravity conjecture and the cosmic censorship conjecture with regard to the thermodynamics of black holes in different structures such as EPST, RPST, and CFT. We aim to challenge these conjectures and the differences in these spaces and examine specific parameters such as AdS radii and central charge etc. Our investigations have revealed that under specific conditions, such as the incorporation of PFDM, these conjectures can indeed coexist harmoniously. This concord is vital as it ensures that our theoretical models remain consistent and do not yield contradictory predictions about the thermodynamic behavior of black holes. By demonstrating compatibility under these conditions, we contribute to a more coherent understanding of the fundamental principles governing black hole physics.
\section{The Model}
Reissner-Nordström AdS black holes, which are enveloped by PFDM, carry an electrical charge and are affected by a theoretical type of matter that constitutes the majority of the universe's mass. These black holes exhibit intriguing thermodynamic behaviors and phase transition characteristics, which can be explored by broadening the phase space of the system. Additionally, some studies have investigated the WCCC in the context of these black holes\cite{601,602,603,604,605,606}. This conjecture posits that a black hole's singularity remains hidden from any external observer. The action under consideration pertains to a gravitational theory that interacts minimally with a gauge field amidst the presence of  PFDM\cite{42,42222,43,44,45},
\begin{equation}\label{eq1}
S=\int d^4x\sqrt{-g}\bigg[\frac{1}{16\pi G}R+\frac{1}{4}F^{\mu\nu}F_{\mu\nu}-\frac{\Lambda}{8\pi G}+\mathcal{L}_{DM}\bigg],
\end{equation}
in the given context, \( g \) represents the determinant of the metric tensor, expressed as \( g = \det(g_{ab}) \), where \( g_{ab} \) are the components of the metric tensor. The scalar curvature is denoted by \( R \), while \( G \) stands for the gravitational constant. The electromagnetic tensor \( F_{\mu\nu} \) is obtained from the gauge potential \( A_{\mu} \). The Lagrangian density \( L_{DM} \) characterizes the  PFDM. Utilizing the principle of least action allows for the derivation of the Einstein field equations, which can be articulated as follows,
\begin{equation}\label{eq2}
\begin{split}
R_{\mu\nu}-\frac{1}{2}g_{\mu\nu}R+\Lambda g_{\mu\nu}=&-8\pi G(T^{M}_{\mu\nu}+T^{DM}_{\mu\nu})\equiv-8\pi GT_{\mu\nu},
\end{split}
\end{equation}
\begin{equation}\label{eq300}
\begin{split}
&F^{\mu\nu}_{;\nu}=0,\\
&F^{\mu\nu;\alpha}+F^{\nu\alpha;\mu}+F^{\alpha\mu;\nu}=0.
\end{split}
\end{equation}
Where the second line of Eq.(\ref{eq300}) is Bianchi identity. The energy-momentum tensor for ordinary matter is symbolized by \( T^M_{\mu\nu} \), and the energy-momentum tensor for PFDM is represented by \( T^{DM}_{\mu\nu} \). These tensors are detailed in\cite{42,42222,43,44,45}. Consequently, the equations can be expressed as follows,
\begin{equation}\label{eq4}
\begin{split}
&T^{\mu}_{\nu}=g^{\mu\nu}T_{\mu\nu},\\
&T^{t}_{t}=-\rho,\hspace{0.5cm}T^{r}_{r}=T^{\theta}_{\theta}=T^{\phi}_{\phi}=P.
\end{split}
\end{equation}
Where $\rho$ and $P$ and energy density and pressure, respectively. In the framework of the RN AdS metric influenced by  PFDM, it is posited that the components of the energy-momentum tensor satisfy \( T^{r}_{r} = T^{\theta}_{\theta} = T^{\phi}_{\phi} = T^{t}_{t} (1 - \delta) \), with \( \delta \) being a constant. This assumption is substantiated by \cite{41,42,43,44,45}. The formulation of the RN AdS metric within the PFDM context is delineated as follows\cite{42222},
\begin{equation}\label{eq5}
\begin{split}
&ds^{2}=-f(r)dt^{2}+f(r)^{-1}dr^{2}+r^{2}(d\theta^{2}+\sin^2\theta d\phi^{2}),\\
&A_{\mu}=(\frac{Q}{r_+},0,0,0),
\end{split}
\end{equation}
where,
\begin{equation}\label{eq6}
f(r)=1-\frac{2MG}{r}+\frac{GQ^2}{r^{2}}-\frac{\Lambda}{3}r^2+\frac{\gamma}{r}\ln(\frac{r}{|\gamma|}).
\end{equation}
In this analysis, the parameters \( M \), \( Q \), and \( \gamma \) signify the mass, charge and the influence of (PFDM) on the black hole, respectively. Also, $\Lambda = -3/\ell^2$
is the cosmological constant with $\ell$ being the AdS radius. In the absence of PFDM (\( \gamma = 0 \)), the space-time metric simplifies to that of a standard RN AdS black hole.
The parameter \( \gamma \) can assume positive values, reflecting different PFDM scenarios. This study contemplates the implications of positive \( \gamma \) on the black hole's characteristics.
Also, the extremality bound for charged black holes in AdS without PFDM, $\gamma = 0$ reads,
\begin{equation*}\label{eq43}
\begin{split}
\frac{|Q|}{\sqrt{G}M}\leq\frac{3|\widetilde{q}|\sqrt{-1+\sqrt{1+2\widetilde{q}^2}}}{-1+2\widetilde{q}^2+\sqrt{1+2\widetilde{q}^2}}\approx \left\{
                \begin{array}{ll}
                  1 & \hbox{$|\widetilde{q}|\ll 1$} \\
                  \frac{3}{(8\widetilde{q}^2)^{1/4}} & \hbox{$|\widetilde{q}^2|\gg 1$,}
                \end{array}
              \right.
\end{split}
\end{equation*}
\begin{equation*}\label{eq43}
\begin{split}
\widetilde{q}^2:=\frac{6GQ^2}{\ell^2}.
\end{split}
\end{equation*}
This nonlinear extremality bound is found by solving $f(r_\star) \leq0$ where $r_\star$ is the largest solution of f'($r_\star$) = 0, where f(r) is given by Eq.(\ref{eq6}) with $\gamma$ = 0.
\section{RPS thermodynamics}
\begin{figure}[h!]
 \begin{center}
 \subfigure[]{
 \includegraphics[height=5cm,width=5.5cm]{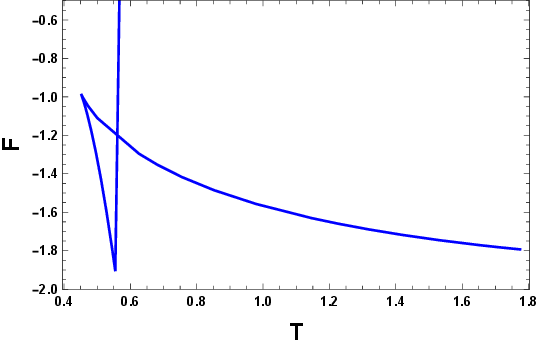}
 \label{fig1a}}
 \subfigure[]{
 \includegraphics[height=5cm,width=5.5cm]{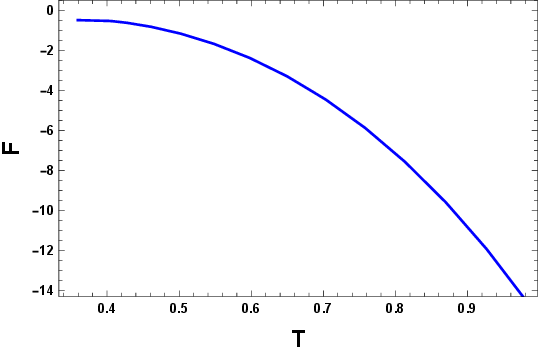}
 \label{fig1b}}
 \subfigure[]{
 \includegraphics[height=5cm,width=5.5cm]{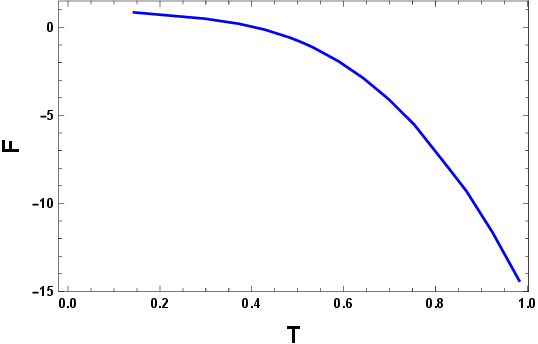}
 \label{fig1c}}
 \caption{\small{With the $\hat{\gamma}=2, C=\ell=1$ and $\hat{Q}_c=0.93$ . (a) $\hat{Q}_c>\hat{Q}=0.5 $ (b) $\hat{Q}_c=\hat{Q}=0.93$ (c)  $\hat{Q}_c<\hat{Q}=1.5$}}.
 \label{fig1}
 \end{center}
 \end{figure}

\begin{figure}[h!]
 \begin{center}
 \subfigure[]{
 \includegraphics[height=5cm,width=5.5cm]{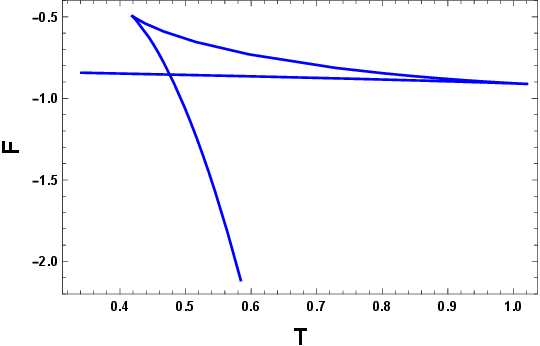}
 \label{fig2a}}
 \subfigure[]{
 \includegraphics[height=5cm,width=5.5cm]{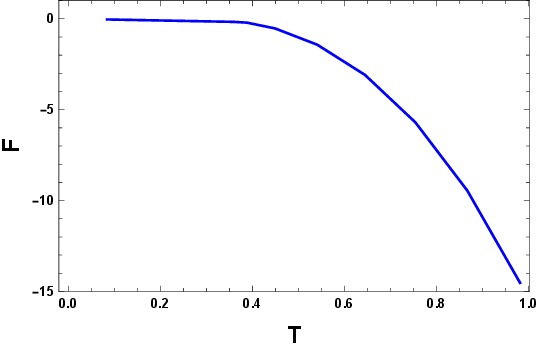}
 \label{fig2b}}
 \subfigure[]{
 \includegraphics[height=5cm,width=5.5cm]{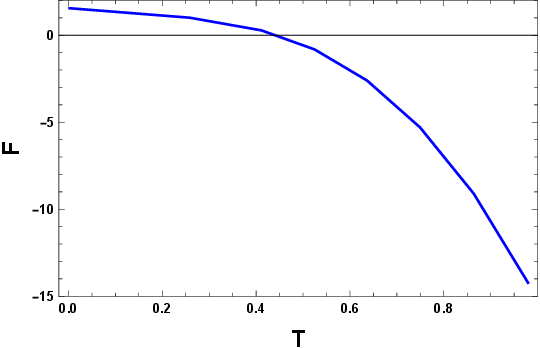}
 \label{fig2c}}
 \caption{\small{With the $\hat{\gamma}=1.5, C=\ell=1$ and $\hat{Q}_c=0.79$ . (a) $\hat{Q}_c>\hat{Q}=0.5 $ (b) $\hat{Q}_c=\hat{Q}=0.79$ (c)  $\hat{Q}_c<\hat{Q}=1.5$}}.
 \label{fig2}
 \end{center}
 \end{figure}

In the RPS formalism, the traditional variables of pressure \( P \) and volume \( V \) are eschewed due to the altered connotations they acquire within the holographic framework. Instead, two pivotal quantities, \( C \) and \( \mu \), assume significant roles within the realm of CFT. The central charge \( C \) is instrumental in ascertaining the microscopic degrees of freedom inherent in CFT, while its conjugate \( \mu \) is regarded as a chemical potential, as referenced in literature\cite{20}. Within the domain of extended phase space thermodynamics, the exploration of black hole thermodynamics necessitates the employment of a dynamic cosmological constant. This adjustment invariably leads to a transformation in the associated dynamical equations. To circumvent this modification, the thermodynamics of constrained phase space is proposed, wherein Newton's constant is treated as a variable, supplanting the cosmological constant. This substitution preserves the original form of the field equations. Moreover, a correlation is established between the central charge and the cosmological and Newton's constants, expressed as \( C = \frac{\ell^2}{ G} \), where \( \ell^2 = -\frac{3}{\Lambda} \) \cite{20}. The first law of thermodynamics, as applied to a RN-AdS black hole surrounded by PFDM within the RPS framework, is articulated as follows,
\begin{equation}\label{eq1'}
dM=TdS+\hat{\phi} d\hat{Q} +\mu dC+\hat{\Pi} d \hat{\gamma},
\end{equation}
where, $\hat{\phi}$ and $\hat{Q}$ are the properly re-scaled electric potential electric charge and $\hat{\Pi}$ is the conjugate of the $\hat{\gamma}$. The observation aligns with an Euler-like relation, which is a fundamental principle in thermodynamics and fluid dynamics, often used to describe the relationship between various state functions. In the context of the RPS formalism, such relations are crucial for understanding the thermodynamic properties of black holes and their behavior in different spacetime geometries. Introducing Euler's relation for RPST can provide a new perspective for studying black hole physics, So,
\begin{equation}\label{eq2'}
M=T S+\hat{\phi}\hat{Q}+\mu C+\hat{\Pi}\hat{\gamma},
\end{equation}
where Eqs.(\ref{eq1'}) and (\ref{eq2'}) are indeed not gauge-invariant at face value. We note that $\hat{\phi}$ represents the difference in electric potential between the black hole horizon and infinity. In the context of our black hole, the parameters \( M \), \( T \), and \( S \) represent the mass, temperature, and entropy, respectively. Moreover, these parameters can generally be characterized by corresponding quantities in the dual CFT,
\begin{equation}\label{eq3'}
\hat{Q}=\frac{Q \ell}{\sqrt{G}} \hspace{1cm}
\hat{\phi}=\frac{\phi\sqrt{G}}{\ell},\hspace{1cm} \hat{\gamma}=\frac{\gamma \ell^2}{G}, \hspace{1cm} \hat{\Pi}=\frac{\Pi G}{\ell^2}.
\end{equation}
We will also have some parameters in RPS as follows,
\begin{equation}\label{eq7}
C=\frac{\ell^2}{G},\qquad    \hat{Q}=Q\sqrt{C} ,\qquad r_+=\ell \sqrt{\frac{S}{\pi  C}},\qquad  \hat{\gamma}=\gamma C.
\end{equation}
Setting $(f(r_+) = 0)$ and utilizing Eqs.(\ref{eq6}), (\ref{eq1'}), (\ref{eq2'}), and (\ref{eq3'}), we can derive some variables such as M, T, $\hat{\phi}$, $\Pi$, and $\mu$ of the black hole as follows,
\begin{equation}\label{eq8}
M=\frac{\pi  C S+\pi ^2 \hat{Q}^2+S^2}{2 \pi ^{3/2} \ell \sqrt{C S}}+\frac{\hat{\gamma}}{2\ell^2}  \ln\bigg(\sqrt{\frac{S C}{\pi  }}\frac{\ell}{\hat{\gamma}}\bigg),
\end{equation}
\begin{equation}\label{eq9}
T=\bigg(  \frac{\partial M}{\partial S} \bigg)_{\hat{Q},C,\hat{\gamma}}=\frac{\pi  C S-\pi ^2 \hat{Q}^2+3 S^2}{4 \pi ^{3/2} \ell S \sqrt{C S}}+\frac{\hat{\gamma} }{4 \ell^2 S}
\end{equation}
\begin{equation}\label{eq10}
\hat{\phi}=\bigg(  \frac{\partial M}{\partial \hat{Q}} \bigg)_{S,C,\hat{\gamma}}= \sqrt{\frac{\pi}{SC}}\frac{\hat{Q} } {\ell },
\end{equation}
\begin{equation}\label{eq11}
\hat{\Pi}=\bigg(  \frac{\partial M}{\partial \hat{\gamma}} \bigg)_{S,\hat{Q},C}=\frac{1}{2\ell^2} \bigg[-1+\ln\bigg(\sqrt{\frac{SC}{\pi  }}\frac{\ell}{\hat{\gamma}}\bigg) \bigg],
\end{equation}
\begin{equation}\label{eq12}
\mu=\bigg(  \frac{\partial M}{\partial C} \bigg)_{S,\hat{Q},\hat{\gamma}}=\frac{\pi C S-\pi ^2 \hat{Q}^2-S^2}{4 \pi ^{3/2} C \ell \sqrt{C S}} +\frac{\hat{\gamma}}{4 C \ell^2}.
\end{equation}
The $T-S$ curve at a fixed value of $\hat{Q}$ and $C$ is a fundamental feature in thermodynamics. It captures the behavior of a system as temperature ($T$) and entropy ($S$) vary. Notably, this curve exhibits a first-order phase transition, which, intriguingly, becomes second-order at the critical point. To identify the critical point on the $T-S$ curve (often referred to as the inflection point), we employ the following equation,
\begin{equation}\label{eq13}
\bigg(\frac{\partial T}{\partial S} \bigg)_{\hat{Q},C,\hat{\gamma}}=0, \qquad   \bigg(\frac{\partial^2 T}{\partial S^2} \bigg)_{\hat{Q},C,\hat{\gamma}}=0.
\end{equation}
Using the above equations for the critical parameters, we have,
\begin{equation}\label{eq14}
Y:=C \ell^{\frac{4}{3}}\big( 81(\frac{\hat{\gamma}}{C})^{2}-\ell^2+9\ell(\frac{\hat{\gamma}}{C})\sqrt{81(\frac{\hat{\gamma}}{C})^{2}-2\ell^2}\big)^{\frac{1}{3}}, \qquad S_c=\frac{\pi (C \ell^2+Y)^2}{18 \ell^2 Y},
\end{equation}
\begin{equation}\label{eq15}
\hat{Q}_c^2=-\frac{\left(C \ell^2+Y\right)^2 \left(C^2 \ell^4-4 C \ell^2 Y+Y^2\right)}{324 \ell^4 Y^2}+\frac{\sqrt{2 }\hat{\gamma}  C^{1/2}  \left(C \ell^2+Y\right)}{9\sqrt{ Y} \ell^2}.
\end{equation}
The critical values of $T$ and $M$ are calculated by,
\begin{equation}\label{eq16}
T_c=\frac{\left(C \ell^2+Y\right) \left(-9 \sqrt{2} C^{1/2} \hat{\gamma}  \ell^2 Y^{3/2}+C^3 \ell^6+6 C^2 \ell^4 Y+6 C \ell^2 Y^2+Y^3\right)}{3  \pi  \sqrt{2C Y} \ell^2 \left(C \ell^2+Y\right)^3}+\frac{9 \hat{\gamma} Y}{2 \pi  \left(C \ell^2+Y\right)^2},
\end{equation}
\begin{equation}\label{eq17}
M_c=\frac{\sqrt{C} \left(C \ell^2+Y\right) \left(2 C \ell^2+3 \hat{\gamma}  \sqrt{\frac{2 Y}{C}}+2 Y\right)}{9 \ell^2 \sqrt{2 Y} \left(C \ell^2+Y\right)}+\frac{\hat{\gamma}}{2 \ell^2}\ln \left(\frac{C(C \ell^2+Y)}{\hat{\gamma}  \sqrt{18 C Y}}\right).
\end{equation}
When studying the thermodynamics of black holes using the Euclidean action, we often encounter a challenge known as redshift degeneracy in the action. To address this issue, we propose considering an appropriate counterterm derived from the AdS/CFT correspondence within the Euclidean action. By doing so, we can resolve this limitation and proceed to find the thermodynamic potential associated with both the canonical and grand canonical ensembles. As an illustrative example, let's refer to the Gibbs energy potential, denoted as G:
$G = I\beta$ where $I$ represent the Euclidean action, and $\beta$ is the inverse of the Hawking temperature. From this Gibbs energy potential, we can derive other essential thermodynamic quantities: Entropy (S): Given by $S =- \frac{dG}{dt}$, angular momentum (J): Given by $J = -\frac{dG}{d\Omega}$, electric charge (Q): Given by $Q =- \frac{dG}{d\phi}$. This approach is equivalent to using the first law of thermodynamics to determine these quantities. Additionally, we can express these thermodynamic quantities in terms of the Helmholtz free energy (F) and its conjugates: Entropy (S): $S = -\frac{dF}{dt}$, angular momentum (J): $J = -\frac{dF}{d\Omega}$, electric charge (Q): $Q = -\frac{dF}{d\phi}$. The connection between the Helmholtz free energy and the Euclidean action is given by $F = I\beta$. Remarkably, the results obtained from the Helmholtz free energy remain consistent with those derived from the first law of thermodynamics. In the RPS formalism, Newton's constant is treated as a variable external to the Euclidean action. Consequently, it does not introduce any additional terms to the counterterm. By adjusting the counterterm appropriately, we can eliminate the redshift degeneracy for different types of black holes. Now, we delve into the Helmholtz free energy,
\begin{equation}\label{eq18}
F=M-TS=\frac{\pi  C S+3 \pi ^2 \hat{Q}^2-S^2}{4 \pi ^{3/2} \ell \sqrt{C S}}+\frac{\hat{\gamma}}{4 \ell^2}\bigg[-1+2 \ln\bigg(\frac{\ell}{\hat{\gamma}}\sqrt{\frac{SC}{\pi}}\bigg)\bigg].
\end{equation}
Figs.(\ref{fig1}) and (\ref{fig2}) depict the evolution of free energy as a function of temperature for various free parameter values of the RN AdS black hole, analyzed within the framework of PFDM in RPS. When \( \hat{Q}_c > \hat{Q} = 0.5 \), a distinctive swallowtail pattern emerges, indicative of a first-order phase transition. Also, for (\( \hat{Q}_c = \hat{Q}  \) ), we have a second-order phase transition. To enhance clarity, we have separated the figures. The specific free parameters pertinent to each figure are duly annotated. The separation of the figures isolates and emphasizes each parameter set's unique characteristics, facilitating a more detailed examination of the system's behavior under varying conditions. It is essential to recognize that the free parameters play a crucial role in shaping the physical properties of the black hole.

\begin{figure}[h!]
 \begin{center}
 \subfigure[]{
 \includegraphics[height=7cm,width=9cm]{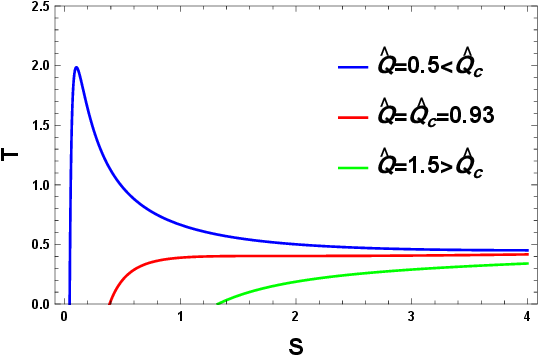}
 \label{fig3a}}
 \subfigure[]{
 \includegraphics[height=7cm,width=9cm]{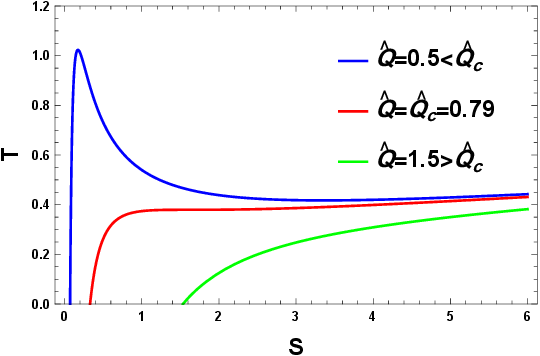}
 \label{fig3b}}
 \caption{\small{spinodal curve ($T-S$) for charged black hole in presence perfect fluid. We have set $C=\ell=1$  and  (a) $\hat{\gamma}=2 $ and $\hat{Q}_c=0.93$, (b)  $\hat{\gamma}=1.5 $ and $\hat{Q}_c=0.79$}}.
 \label{fig3}
 \end{center}
 \end{figure}
Fig. (\ref{fig3}) presents the Hawking temperature as a function of entropy for the AdS (RN) black hole amidst the presence of (PFDM) in the (RPS). We have set the constants \( C \) and $\ell$ to 1. For Fig.(\ref{3a}), the parameters are set to \( \hat{\gamma} = 2 \) and $\hat{Q}_c$= 0.93 , while for Figure 3b, \( \hat{\gamma} = 1.5 \) and $\hat{Q}_c$= 0.79 . Each subplot within the figure reveals certain zero points for varying free parameters within the range \( 0 < S < 1 \). Upon examining the entropy alongside the black hole's free parameter values, it is observed that for small values of \( S \), the structural behavior of the black hole is markedly unique. Conversely, as the value of \( S \) increases, the figures tend to converge, indicating a unification in behavior for larger entropy values. This convergence suggests that the influence of the free parameters diminishes as the entropy grows, leading to a more predictable and uniform structure for the black hole.
\section{CFT thermodynamics}
\begin{figure}[h!]
 \begin{center}
 \subfigure[]{
 \includegraphics[height=5cm,width=5.5cm]{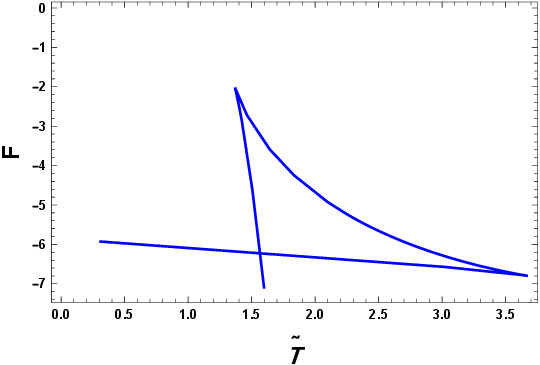}
 \label{fig4a}}
 \subfigure[]{
 \includegraphics[height=5cm,width=5.5cm]{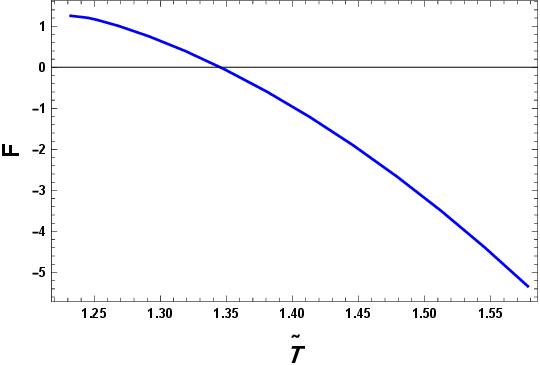}
 \label{fig4b}}
 \subfigure[]{
 \includegraphics[height=5cm,width=5.5cm]{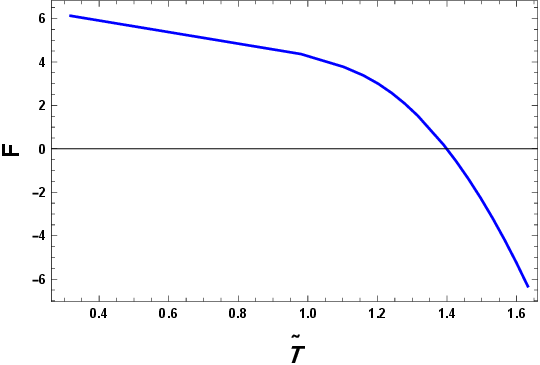}
 \label{fig4c}}
 \caption{\small{With the $\beta=2, C=\mathcal{V}=1$ and $\tilde{Q}_c=2.52$ . (a) $\tilde{Q}_c>\tilde{Q}=1.5 $ (b) $\tilde{Q}_c=\tilde{Q}=2.52$ (c)  $\tilde{Q}_c<\tilde{Q}=3$}}.
 \label{fig4}
 \end{center}
 \end{figure}

\begin{figure}[h!]
 \begin{center}
 \subfigure[]{
 \includegraphics[height=5cm,width=5.5cm]{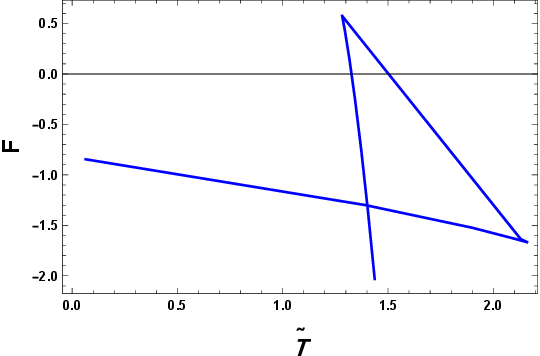}
 \label{fig5a}}
 \subfigure[]{
 \includegraphics[height=5cm,width=5.5cm]{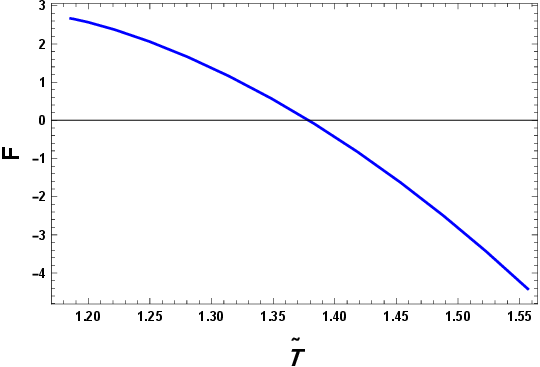}
 \label{fig5b}}
 \subfigure[]{
 \includegraphics[height=5cm,width=5.5cm]{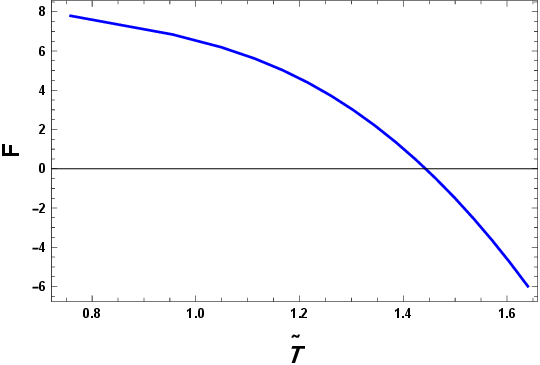}
 \label{fig5c}}
 \caption{\small{With the $\beta=1.5, C=\mathcal{V}=1$ and $\tilde{Q}_c=2.15$ . (a) $\tilde{Q}_c>\tilde{Q}=1.5 $ (b) $\tilde{Q}_c=\tilde{Q}=2.15$ (c)  $\tilde{Q}_c<\tilde{Q}=3$}}.
 \label{fig5}
 \end{center}
 \end{figure}

To explore the thermodynamics of the charged AdS black hole with PFDM within the CFT framework, it's essential to understand the holographic links that connect the bulk and boundary quantities. We denote the boundary's curvature radius by \( R \), which is distinct from the AdS radius \( \ell \) in the bulk. The CFT metric, showcasing conformal scaling invariance, is given by,
\begin{equation}\label{eq41}
\begin{split}
ds^2=\omega^2(-dt^2+\ell^2 d\Omega_{d-2}^2).
\end{split}
\end{equation}
Here, \( \omega \) is the dimensionless conformal factor that varies freely, signifying the boundary theory's conformal symmetry. In a spherical scenario, \( d\Omega_{d-2}^2 \) is the metric on a \( (d-2) \)-dimensional sphere. We assume \( \omega \) is constant across the \begin{equation}\label{eq42}
\begin{split}
\mathcal{V}=\Omega_{d-2} R^{d-2},
\end{split}
\end{equation}
where \( R = \omega \ell \) signifies the manifold's variable curvature radius where the CFT resides. Notably, variations in the central charge \( C \) do not affect \( \mathcal{V} \). In Einstein's gravity context, \( C \) and \( \ell \) exhibit a dual relationship, as indicated in a previous equation. We maintain a constant Newton's constant \( G \) while allowing \( \ell \) to vary, which in turn alters \( C \). The holographic dictionary facilitates the correlation of bulk parameters \( M, T, S, \Phi, Q \) with their CFT counterparts \( E, \tilde{T}, \tilde{S}, \tilde{\Phi}, \tilde{Q} \), as follows,
\begin{equation}\label{eq43}
\begin{split}
E=\frac{M}{\omega}, \qquad  \tilde{T}=\frac{T}{\omega}, \qquad  \tilde{S}=S, \qquad \tilde{Q}=\frac{Q \ell}{\sqrt{G}}, \qquad  \tilde{\Phi}=\frac{\Phi \sqrt{G}}{\omega \ell}.
\end{split}
\end{equation}
Employing these relationships, we can reformulate the first law of thermodynamics for the charged AdS black hole in the CFT context:
\begin{equation}\label{eq44}
\begin{split}
\delta E= \tilde{\Phi} \delta \tilde{Q}+ \mu \delta C-p\delta \mathcal{V} +\tilde{T} \delta S,
\end{split}
\end{equation}
where the chemical potential \( \mu \) and pressure \( p \) are defined as,
\begin{equation}\label{eq45}
\begin{split}
\mu=\frac{E-\tilde{T}S-\tilde{\Phi} \tilde{Q}}{C},
\end{split}
\end{equation}
\begin{equation}\label{eq46}
\begin{split}
p=\frac{E}{(d-2)\mathcal{V}}.
\end{split}
\end{equation}
Furthermore, it's evident from the above equation that \( \mathcal{V} \) and \( C \) vary independently. This paper aims to delve into the consequences of this theoretical framework on the thermodynamics of RN AdS black holes in the presence of PFDM within the CFT paradigm. Our objective is to reformulate the bulk thermodynamic first law in a manner that incorporates the boundary central charge. To achieve this, we turn to the holographic dual relationship between key quantities in Einstein's gravity and the AdS scale, the central charge C, and Newton's constant G. In this context, we consider a setting where both the cosmological constant ($\Lambda$) and gravitational Newton constant ($G$) are allowed to vary within the bulk. By doing so, we can express the first law in a new form that involves both $\Lambda$ (associated with thermodynamic pressure) and the central charge $C$ from the dual CFT. These variables and their conjugates play a crucial role in understanding the thermodynamics of AdS black holes. This approach allows us to explore novel aspects of black hole behavior, including phase transitions, critical points, and universal scaling behavior, all while maintaining a fixed boundary central charge. The interplay between holography and bulk thermodynamics provides valuable insights into the nature of quantum gravity,
\begin{equation}\label{eq19}
C=\frac{\Omega_{d-2}\ell^{d-2}}{16\pi G}.
\end{equation}
Here, we aim to derive the first law of thermodynamics within the context of the CFT mode for the RN-AdS black hole with PFDM. Our goal is to establish a meaningful correspondence between quantities in the bulk (associated with the black hole) and those on the boundary (related to the CFT). To achieve this, we introduce a scale transformation given by:$E = \frac{M}{\omega}$ where $\omega = \frac{R}{\ell}$ represents the ratio of the AdS radius ($R$) to the characteristic length scale ($\ell$). We focus on the case where the spacetime dimension is $d=4$. By analyzing the modified equations, we explore the interplay between the RN-AdS black hole, its thermodynamic properties, and the dual CFT description. This investigation sheds light on the intricate relationship between the bulk geometry and the boundary theory, providing valuable insights into the underlying physics,
\begin{equation}\label{eq20}
\Omega_{2}=4\pi, \qquad  C=\frac{\ell^{2}}{4 G},
\end{equation}
and
\begin{equation}\label{eq21}
\begin{split}
& \delta\bigg( \frac{M}{\omega}\bigg)= \frac{T}{\omega} \delta \bigg(\frac{A}{4G} \bigg)+\bigg(\frac{M}{\omega}-\frac{TS}{\omega}-\frac{Q\Phi}{\omega}-\frac{\Pi\gamma}{\omega} \bigg) \frac{\delta ( \ell^{2}/4G)}{ \ell^{2}/4G}\\
&-\frac{M}{2\omega}\frac{\delta (\Omega_2 R^{2})}{(\Omega_2 R^{2})}+\frac{\Phi \sqrt{G}}{\omega\ell} \delta \bigg(\frac{Q\ell}{\sqrt{G}} \bigg)+\frac{\Pi \sqrt{G}}{\omega\ell} \delta \bigg(\frac{\gamma\ell}{\sqrt{G}} \bigg).
\end{split}
\end{equation}
Based on the equation provided above, the correspondence between bulk and boundary quantities for the black hole can be established as follows,
\begin{equation}\label{eq22}
\begin{split}
\delta E= \tilde{T} \delta \tilde{S}+\mu \delta C-p \delta \mathcal{V}+ \tilde{\Phi} \delta \tilde{Q}+  \tilde{\Pi} \delta \tilde{\gamma},
\end{split}
\end{equation}
where
\begin{equation}\label{eq23}
\begin{split}
E=\frac{M}{\omega}, \quad  \tilde{S}=S,\quad \tilde{T}=\frac{T}{\omega}, \quad  \tilde{\Phi}=\frac{\Phi\sqrt{G}}{\omega\ell},\quad \tilde{Q}=\frac{Q\ell}{\sqrt{G}}, \quad \tilde{\Pi}=\frac{\Pi\sqrt{G}}{\omega \ell},\quad \tilde{\gamma}=\frac{\gamma\ell}{\sqrt{G}},
\end{split}
\end{equation}
if $\beta=\frac{\tilde{\gamma}}{\ell}$, we also calculate thermodynamic quantities for the CFT as follows,
\begin{equation}\label{eq24}
\begin{split}
E=\frac{4 \pi  C S+\pi ^2 \tilde{Q}^2+S^2}{2 \pi  \sqrt{C S \mathcal{V}}}+\frac{2 \beta  \sqrt{\pi  C}}{\sqrt{\mathcal{V}}}\ln \left(\frac{\sqrt{S}}{\sqrt{\pi } \beta }\right),
\end{split}
\end{equation}

\begin{equation}\label{eq25}
\tilde{T}=\bigg(  \frac{\partial E}{\partial S} \bigg)_{\tilde{Q},C,\beta,\mathcal{V}}=\frac{4 \pi  C S-\pi ^2 \tilde{Q}^2+3 S^2}{4 \pi  S^{3/2} \sqrt{C \mathcal{V}}}+\frac{\beta  \sqrt{\pi  C}}{S \sqrt{\mathcal{V}}},
\end{equation}

\begin{equation}\label{eq26}
\tilde{\Phi}=\bigg(  \frac{\partial E}{\partial \tilde{Q}} \bigg)_{C,S,\beta,\mathcal{V}}=\frac{\pi  \tilde{Q}}{\sqrt{C S \mathcal{V}}},
\end{equation}

\begin{equation}\label{eq27}
\mu=\bigg(  \frac{\partial E}{\partial C} \bigg)_{\tilde{Q},S,\beta,\mathcal{V}}=\frac{4 \pi  C S-\pi ^2 \tilde{Q}^2-S^2}{4 \pi C \sqrt{C S \mathcal{V}}}+\frac{\sqrt{\pi } \beta }{\sqrt{C \mathcal{V}}}\ln\bigg(\frac{\sqrt{S}}{\sqrt{\pi } \beta } \bigg),
\end{equation}

\begin{equation}\label{eq28}
\tilde{\Pi}=\bigg(  \frac{\partial E}{\partial \tilde{\gamma}} \bigg)_{\tilde{Q},S,C,\mathcal{V}}=\frac{\beta  \sqrt{\pi  C}}{\tilde{\gamma} S \sqrt{\mathcal{V}}},
\end{equation}

\begin{equation}\label{eq29}
-p=\bigg(  \frac{\partial E}{\partial \mathcal{V}} \bigg)_{\tilde{Q},S,C,\beta}=-\frac{E}{2\mathcal{V}}.
\end{equation}
Additionally, we can utilize the following relationships to identify the critical points,
\begin{equation}\label{eq30'}
\begin{split}
\bigg(\frac{\partial \tilde{T}}{ \partial S} \bigg)_{\tilde{Q},\mathcal{V},C,\beta}=0, \qquad   \bigg(\frac{\partial^2 \tilde{T}}{ \partial S^2} \bigg)_{\tilde{Q},\mathcal{V},C,\beta}=0.
\end{split}
\end{equation}
By applying the connections between equations \(\eqref{eq25}\) and \(\eqref{eq30'}\), we can determine the critical points as outlined below
\begin{equation}\label{eq30}
X:=(-4 C^3+81 \beta ^2 C^2+9\beta C^2 \sqrt{81 \beta ^2 -8  C})^{\frac{1}{3}}, \qquad S_c=\frac{1}{9} \left(\frac{ 2^{5/3} \pi  C^2}{X}+4 \pi  C+2^{\frac{1}{3}} \pi  X\right),
\end{equation}

\begin{equation}\label{eq31}
\tilde{Q}_c^2=\frac{-2^{10/3} C^4+2^{11/3} C^3 X+24 C^2 X^2+2^{7/3} C X^3-2^{2/3} X^4}{81 X^2}+\frac{8}{9} \beta  C \sqrt{\frac{2^{5/3} C^2}{X}+4 C+2^{\frac{1}{3}} X},
\end{equation}

\begin{equation}\label{eq32}
\tilde{T}_c=\frac{4 \pi  C S_c-\pi ^2 \tilde{Q}_c^2+3 S_c^2}{4 \pi  S_c^{3/2} \sqrt{C \mathcal{V}}}+\frac{\beta  \sqrt{\pi  C}}{S_c \sqrt{\mathcal{V}}}, \qquad
E_c=\frac{4 \pi  C S_c+\pi ^2 \tilde{Q}_c^2+S_c^2}{2 \pi  \sqrt{C S_c \mathcal{V}}}+\frac{2 \beta  \sqrt{\pi  C}}{\sqrt{\mathcal{V}}}\ln \bigg(\frac{\sqrt{S_c}}{\sqrt{\pi } \beta }\bigg).
\end{equation}
We utilize equations  \(\eqref{eq24}\) and  \(\eqref{eq25}\) to determine the Helmholtz energy.
\begin{equation}\label{eq33}
F=E-\tilde{T} \tilde{S}=\frac{12 \pi  C S+\pi ^2 \tilde{Q}^2+5 S^2}{4 \pi  \sqrt{C S\mathcal{V}}}+\frac{\beta  \sqrt{\pi  C}}{\sqrt{\mathcal{V}}}\bigg[-1+2 \ln \bigg(\frac{\sqrt{S}}{\sqrt{\pi } \beta }\bigg)\bigg].
\end{equation}
Figs. (\ref{fig4}) and (\ref{fig5}) showcase the evolution of free energy as a function of temperature for diverse free parameter values of the RN AdS black hole, situated within the theoretical framework of PFDM in Conformal Field Theory. A notable swallowtail pattern is observed when \( \tilde{Q}_c > \tilde{Q} = 1.5 \), signaling a first-order phase transition. Also, in $\tilde{Q}_c=\tilde{Q}$, we face with second phase transition. For enhanced visual clarity, we have separated the figures. Each figure is meticulously labeled with the relevant free parameters. By separating the figures, we aim to highlight and scrutinize the distinct features of each parameter set, thereby enabling a more granular analysis of the system's dynamics under different conditions. The role of free parameters is pivotal in determining the physical attributes of the black hole.

\begin{figure}[h!]
 \begin{center}
 \subfigure[]{
 \includegraphics[height=7cm,width=9cm]{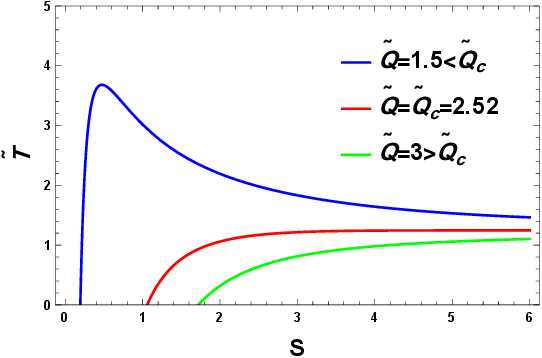}
 \label{fig6a}}
 \subfigure[]{
 \includegraphics[height=7cm,width=9cm]{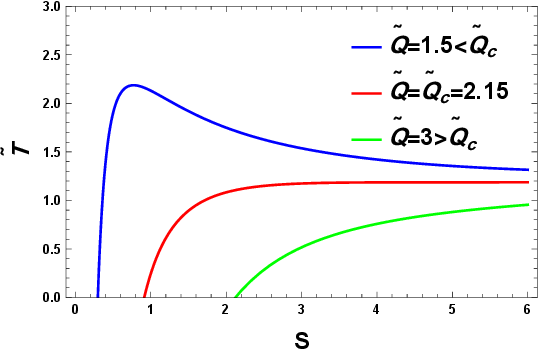}
 \label{fig6b}}
 \caption{\small{spinodal curve ($\tilde{T}-S$) for charged black hole in presence perfect fluid. We have set $C=\mathcal{V}=1$  and  (a) $\beta=2 $ and $\tilde{Q}_c=2.52$, (b)  $\beta=1.5 $ and $\tilde{Q}_c=2.15$}}.
 \label{fig6}
 \end{center}
 \end{figure}

Fig.(\ref{fig6}) illustrates the Hawking temperature as a function of entropy for the RN AdS black hole in the context of  PFDM within CFT. We set \( C \) =\( \mathcal{V} \)=1. For Figure 6a, the parameters are chosen as \( \beta = 2 \) and \( \tilde{Q}_c = 2.52 \), whereas for Fig. (\ref{6b}), \( \beta = 1.5 \) and \( \tilde{Q}_c = 2.15 \) are selected. Each subplot in the figure discloses distinct zero points for varying free parameters across the entropy range \( 0 < S < 1 \). A detailed analysis of the entropy of the black hole's free parameter values reveals a pronounced uniqueness in structural behavior for lower entropy values, denoted by \( S \). This uniqueness is characterized by significant deviations in the thermal properties of the black hole. In contrast, as the entropy value \( S \) ascends, the subplots tend to align, signifying a behavioral unification at higher entropy levels.
The potential violation of the second law of thermodynamics in the context of black hole thermodynamics has been a topic of significant interest, particularly in the various frameworks.
In EPST, the cosmological constant is treated as a thermodynamic variable analogous to pressure, and its conjugate variable is the volume. This framework allows for a richer thermodynamic description of black holes, including phase transitions similar to those in conventional thermodynamics.
RPST is a newer framework that excludes the \(PdV\) term from the first law of black hole thermodynamics, focusing instead on other thermodynamic variables such as the central charge and its conjugate chemical potential.
In the context of the AdS/CFT correspondence, black hole thermodynamics can be studied using the dual CFT. This approach provides a powerful tool for understanding the microscopic origins of black hole entropy and the second law.
in\cite{666a,666b,666c} researchers have observed violations of the second law under certain conditions, particularly for extremal and near-extremal black holes. These violations are often linked to the behavior of the black hole's entropy and the thermodynamic volume.
More recent work\cite{666',666''}, suggests that these violations can be resolved by considering additional thermodynamic variables or by imposing certain constraints on the system. For instance, the inclusion of quantum corrections or the consideration of higher-dimensional spacetimes can restore the validity of the second law.
Also shows potential violations of the second law, particularly in the context of charged black holes. These violations are often associated with the specific thermodynamic paths taken by the system.
The researchers indicate that by carefully choosing the thermodynamic variables and paths, the second law can still hold. This involves ensuring that the entropy production is always non-negative, even in the presence of complex interactions and phase transitions.
Recent works suggest that by considering the full spectrum of states in the CFT and including quantum corrections, the second law can be upheld. This involves ensuring that the entropy of the black hole and the dual CFT system always increases, even under extreme conditions.
The potential violation of the second law of thermodynamics in black hole systems is a complex issue that depends heavily on the chosen thermodynamic framework. While violations have been observed under certain conditions, recent research indicates that these can often be resolved by considering additional variables, quantum corrections, or specific constraints. The frameworks of EPST, RPST, and CFT phase spaces each offer unique insights and mechanisms for ensuring the validity of the second law. Therefore, we consider an additional variable for the black hole, referred to as PFDM. This variable not only resolves the conflict between the WGC and the WCCC but also addresses this violation.
\newpage
\section{WGC and WCCC}
The primary objective of this work is to challenge the WGC and WCCC in different spaces. So first we investigate the thermodynamic properties of Reissner-Nordström black holes in the presence PFDM, utilizing the principles of RPS and CFT thermodynamics. Then we explore the special ranges where these conjectures are compatible with each other in the $Q/M>\sqrt{G}$. Our goal is to delve into some crucial concepts. This research is designed to enhance our comprehension of black hole physics, particularly in the context of dark matter and dark energy. We also study quantum gravity viz relation between quantum mechanics and cosmology through WGC and WCCC. So, it allows for an in-depth examination of the interlinked theories and their impact on the thermodynamics and stability of black holes.
\begin{figure}[h!]
 \begin{center}
 \subfigure[]{
 \includegraphics[height=5.5cm,width=5.5cm]{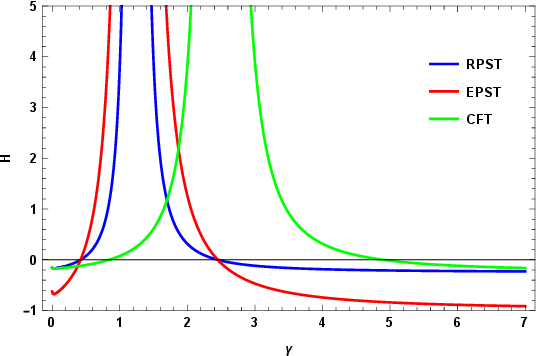}
 \label{fig7a}}
 \subfigure[]{
 \includegraphics[height=5.5cm,width=5.5cm]{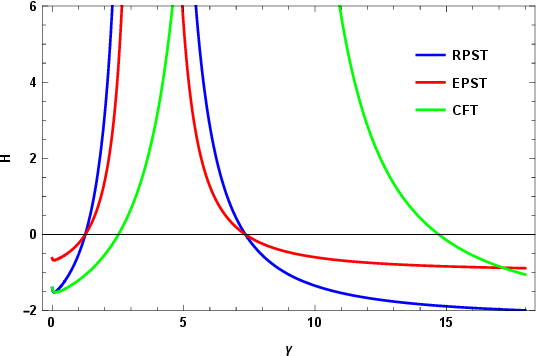}
 \label{fig7b}}
 \subfigure[]{
 \includegraphics[height=5.5cm,width=5.5cm]{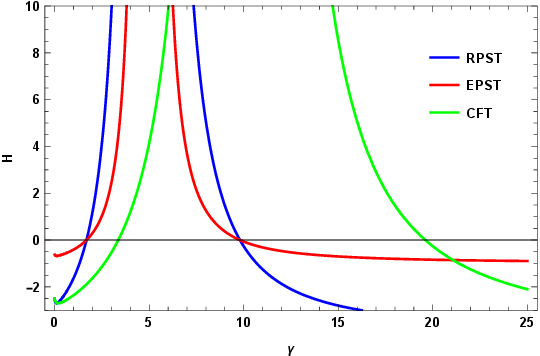}
 \label{fig7c}}
 \caption{\small{Compatibility range of the weak gravity conjecture with different thermodynamics (EPST, RPST, CFT) according to free parameters. (a) $\ell=0.5 $, \hspace{1cm} (b) $\ell=1.5 $,\hspace{1cm} (c)  $\ell=2 $}}.
 \label{fig7}
 \end{center}
 \end{figure}

To substantiate the WGC within various phase spaces, we must establish certain specific relations. In the EPST, the condition is \( H = \frac{Q_c^2}{M_c^2} - 1 > 0 \), and for the RPST and CFT, the criterion is \( H = \frac{Q_c^2}{M_c^2} - G > 0 \).\footnote{We have that $\frac{Q^2}{M^2} > \frac{1}{M_{\text{pl}}^2}$, where $M_{\text{pl}} = \sqrt{\frac{hc}{G}}$, in the extended phase space (where $h=c=G=1$), this inequality holds in Planck units. Therefore, we will have $\frac{Q^2}{M^2} > 1$. Also in RPST, we will have $\frac{Q^2}{M^2} > G$. We can rewrite these terms into this form for convenience with function $H$. Thus, we used the helper function $H$ described above better to represent the WGC in the figures.} Consequently, we can pursue several approaches, among which thermodynamics and the critical values of black hole parameters are particularly significant. In light of this, we utilize critical points obtained from Eq.(\ref{eq30'}) for specific critical quantities in various spaces to verify the Weak Gravity Conjecture. Given that the critical values related with two parameters, namely \( \gamma \) and $\ell$, it becomes evident that these parameters play pivotal roles in validating the WGC. As depicted in Fig. (\ref{fig7}), the ranges within which the  WGC is satisfied—specifically (\( H > 0 \)) for EPST and RPST are the same. However, for CFT, this range is different and larger than other spaces. In relation to the more consistent WGC with CFT, several valuable points can be highlighted. As stated in this section, based on the calculations and the acceptable parameter range for the three thermodynamic states (EPST, RPST, and CFT), it is clear that the CFT section has a higher compatibility range than the other two conjectures. This indicates that in CFT thermodynamics, we can consider a broader compatibility interval compared to the other two states.
A particularly noteworthy point is that since both CFT and WGC are derived from string theory, they may exhibit greater harmony and compatibility with each other. This inherent connection suggests that the principles governing CFT and WGC might be more aligned, leading to a more unified theoretical framework. The implications of this are profound, as it could pave the way for new insights into the fundamental nature of our universe.
It appears that black holes and CFT provide an excellent playground for exploring the weak gravity conjecture. Black holes, with their extreme gravitational fields and unique thermodynamic properties, offer a rich testing ground for theoretical predictions. The interplay between black hole physics and CFT could reveal deeper connections between gravity and quantum field theories, potentially leading to breakthroughs in our understanding of both.
Our research has focused on examining the conditions under which the WGC holds true in various theoretical models, including those involving black holes and CFT. By analyzing these models, we aim to uncover patterns and relationships that could provide further evidence supporting the WGC.
Moreover, we are seeking more evidence to support this hypothesis as well as string theory. This involves not only theoretical investigations but also potential experimental or observational tests that could validate our predictions. The quest for such evidence is ongoing, and each new discovery brings us closer to a more comprehensive understanding of the fundamental forces that govern our universe.
In summary, the relationship between the Weak Gravity Conjecture and Conformal Field Theory is a promising area of research. The higher compatibility range of CFT thermodynamics, the potential harmony between CFT and WGC due to their common origin in string theory, and the rich testing ground provided by black holes all contribute to this exciting field of study. Our ongoing work aims to further explore these connections and gather more evidence to support our hypotheses.

$$ RPST=EPST \rightarrow \sqrt{\frac{2}{3}}\ell<\gamma<5\ell$$,

$$ CFT \rightarrow \sqrt{3}\ell<\gamma<10\ell$$.

Furthermore, through computational methods, we have approximated these intervals. Additionally, upon scrutinizing the WCCC at the critical juncture, we ascertain that it remains valid viz $f(M_c, Q_c, r_c)=0$. Thus, it can be concluded that even though the WGC is upheld for the black hole at the critical point, the WCCC is also concurrently sustained. This dual validity underscores the robustness of the conjectures within the critical framework of black hole parameters.\\\\
As can be seen above the  WGC in CFT thermodynamic space yields a better result and is within a more acceptable range. The coordination of the WGC and CFT can be checked in different works\cite{455,4555,45555,455555}.\\
The presence of PFDM plays a pivotal role in reconciling the WGC and the WCCC within the context of RN black holes. We analyze how PFDM helps resolve the conflict between these two conjectures.\\
WGC posits that gravity should be the weakest force in any consistent theory of quantum gravity. It implies that there must be a particle with a charge-to-mass ratio greater than or equal to that of an extremal black hole, ensuring that black holes can decay.
Also WCCC asserts that singularities arising from gravitational collapse are hidden within event horizons, preventing them from being observed from the rest of spacetime. This ensures the predictability of physical laws.
PFDM modifies the spacetime geometry around RN black holes, which in turn affects the conditions under which the WGC and WCCC are satisfied. In the absence of PFDM, an RN black hole can either have two event horizons (if \( Q^2/M^2 \leq 1 \)) or none (if \( Q^2/M^2 > 1 \)). The latter scenario results in a naked singularity, violating the WCCC.
When PFDM is present, it alters the metric of the black hole, ensuring that the singularity is always covered by an event horizon, thus satisfying the WCCC. This is achieved by introducing a parameter \( \gamma \) which represents the influence of PFDM.\\
PFDM changes the conditions for the existence of event horizons. By adjusting the parameter \( \gamma \), the RN black hole can maintain its event horizons even when \( Q^2/M^2 > 1 \), preventing the formation of a naked singularity and thus upholding the WCCC. There exists a critical value of \( \gamma \), denoted as \( \gamma_{ext} \), which makes the RN black hole extremal. At this value, the black hole has an event horizon, ensuring that both the WGC and WCCC are satisfied. The parameter \( \gamma \) and the radius of AdS space \( l \) have a linear relationship with the range compatibility of the WGC. This means that the presence of PFDM expands the range of parameters for which the WGC is satisfied, making it more consistent with the WCCC.
PFDM effectively reconciles the WGC and WCCC by modifying the spacetime geometry around RN black holes, ensuring that singularities are always hidden within event horizons and that gravity remains the weakest force. This dual satisfaction of the conjectures demonstrates the robustness of these theoretical frameworks when PFDM is considered.

In summary, the WGC in CFT thermodynamic space is favored because it integrates well with established principles of quantum field theory and quantum gravity, offers a stable thermodynamic interpretation, and avoids theoretical inconsistencies, making it a more reliable and acceptable result within the theoretical physics community.
\section{Conclusions}
In this paper, we explore the intriguing interplay between fundamental theoretical physics concepts within the context of charged black holes. Specifically, we focus on the consistency of the  WGC and WCCC in CFT, and  RPS for RN AdS-PFDM. First, we have meticulously explored the thermodynamic properties of RN AdS black holes in the presence of  PFDM within the realm of RPS and CFT. Our research has been underpinned by a series of detailed figures that depict the evolution of free energy and Hawking temperature as functions of temperature and entropy, respectively. Our findings reveal a remarkable swallowtail pattern in the free energy of RN AdS black holes, which emerges under specific free parameter values and signifies a first-order phase transition. We found that despite the differences in EPST, RPST and CFT spaces, similar phase behavior is observed in all these spaces. In our study, we delve into the intricacies of the " WGC", which is a theoretical framework within the field of quantum gravity. The WGC posits that for any gauge force to be consistent with quantum gravity, it must exert forces that are stronger than gravitational forces on some particles. This conjecture has profound implications across various phase spaces, and we specifically focus on its application in "EPST", " RPST, and CFT" spaces. Our investigation highlights the critical role that thermodynamic principles and black hole parameters play in substantiating the WGC. Through rigorous exploration, we delineate the regions within these phase spaces where the WGC holds met. Interestingly, we observe that the parameter space accommodating the WGC is more expansive in the context of CFT, surpassing the confines observed in EPST and RPST spaces. Moreover, we pinpoint specific intervals for the parameters \( \gamma \) and \( l \) that lend credence to the WGC. By employing computational techniques, we not only bolster the WGC but also affirm the "WCCC". The WCCC, which is a fundamental principle in the study of black holes, asserts that singularities arising within black holes should be concealed from external observers, thereby preserving the predictability of physical laws. Hence, for a charged black hole, when $(Q/M>\sqrt{G})$, we will face a naked singularity, which violates WCCC, and therefore is not compatible with WGC, but in this case, adding PFDM will solve the mentioned problem and the compatibility of these two conjectures appears. In a pivotal examination at the critical juncture of black hole parameters, we scrutinize the WCCC and establish its validity. The critical condition \( f(M_c, Q_c, r_c) = 0 \) serves as a testament to the conjecture's soundness. Our findings lead us to conclude that both the WGC and WCCC receive substantial support at the critical points of black holes. This not only demonstrates the robustness of these conjectures but also underscores their significance within the critical areas of black hole physics.

\end{document}